# Towards the 1G of Mobile Power Network: RF, Signal and System Designs to Make Smart Objects Autonomous

**Bruno Clerckx[1], Alessandra Costanzo[2], Apostolos Georgiadis[3], and Nuno Borges Carvalho[4]**

[1]Imperial College London, UK, [2]University of Bologna, Italy, [3]Heriot-Watt University, UK, [4]University of Aveiro, Portugal

Email: b.clerckx@imperial.ac.uk, alessandra.costanzo@unibo.it, A.Georgiadis@hw.ac.uk, nbcarvalho@ua.pt

Thanks to the quality of the technology and the existence of international standards, wireless communication networks (based on radio-frequency RF radiation) nowadays underpin the global functioning of our societies. The pursuit towards higher spectral efficiency has been around for about 4 decades, with 5G expected in 2020. 5G and beyond will see the emergence of trillions of low-power autonomous wireless devices for applications such as ubiquitous sensing through an Internet of Things (IoT).

Wireless is however more than just communications. For very short range, wireless power via Inductive Power Transfer is a reality with available products and standards (Wireless Power Consortium, Power Matters Alliance, Alliance for Wireless Power, Rezence). Wireless Power via RF (as in wireless communication) on the other hand could be used for longer range via two different ways, commonly referred to as wireless energy harvesting (WEH) and (far-field or radiative) wireless power transfer/transmission (WPT). While WEH assumes RF transmitters are exclusively designed for communication purposes whose ambient signals can be harvested, WPT relies on dedicated sources designed exclusively for wireless power delivery. Wireless Power via RF has long been regarded as a possibility for energising low-power devices, but it is only recently that it has become recognised as feasible. Indeed, according to [Hemour:2014], at a fixed computing load, the amount of requested energy falls by a factor of two every year and a half due to the evolution of the electrical efficiency of computer technology. This explains why relying on wireless power to perform meaningful computation tasks at reasonable distances only became feasible in the last few years and justifies this recent interest in wireless power.

Recent research advocates that the future of wireless networking goes beyond conventional communication-centric transmission. In the same way as wireless (via RF) has disrupted mobile communications for the last 40 years, wireless (via RF) will disrupt the delivery of mobile power. However, current wireless networks have been designed for communication purposes only. While mobile communication has become a relatively mature technology, currently evolving towards its fifth generation, the development of mobile power is in its infancy and has not even reached its first generation. Not a single standard on mobile power and far-field WPT exists.

Despite being subject to regulations on exposure to electromagnetic fields as wireless communication, wireless power brings numerous new opportunities. It enables proactive and controllable energy replenishment of devices for genuine mobility so that they no longer depend on centralised power sources. Hence, no wires, no contact, no (or at least reduced) batteries (and therefore smaller, lighter and compact devices), an ecological solution with no production/maintenance/disposal of trillions of batteries, a prolonged lifetime and a perpetual, predictable and reliable energy supply as opposed to ambient energy-harvesting technologies (solar, thermal, vibration). This is very relevant in future networks with ubiquitous and autonomous low-power and energy limited devices, device-to-device communications and the Internet-of-Things (IoT) with massive connections.

Interestingly, radio waves carry both energy and information simultaneously. Nevertheless, traditionally, energy and information have been treated separately and have evolved as two independent fields in academia and industry, namely wireless power and wireless communication, respectively. This separation has for consequences that 1) current wireless networks pump RF energy into the free space (for communication purposes) but do not make use of it for energizing devices and 2) providing ubiquitous mobile power would require the deployment of a separate network of dedicated energy transmitters. Imagine instead a wireless network where information and energy flow together through the wireless medium. Wireless communication, or Wireless Information Transfer (WIT), and WPT would refer to two extreme strategies respectively targeting communication-only and power-only. A unified Wireless Information and Power Transfer (WIPT) design would have the ability to softly evolve in between those two extremes to make the best use of the RF spectrum/radiations and network infrastructure to communicate and energize, and hence outperform traditional systems relying on a separation of communications and power.

This article reviews some recent promising approaches to make the above vision closer to reality. In contrast with articles commonly published by the microwave community and the communication/signal processing community that separately emphasize RF, circuit and antenna solutions for WPT on one hand and communications, signal and system designs for WPT on the other hand, this review article uniquely bridges RF, signal and system designs in order to bring those communities closer to each other and get a better understanding of the fundamental building blocks of an efficient WPT network architecture. We start by reviewing the engineering requirements and design challenges of making mobile power a reality. We then review the state-of-the-art in a wide range of areas spanning sensors and devices, RF design for wireless power and wireless communications. We identify their limitations and make critical observations before providing some fresh new look and promising avenues on signal and system designs for WPT.

**Engineering Requirements and Design Challenges of the Envisioned Network**
The followings are believed to be the engineering requirements and the main design challenges: *1) Range:* Deliver wireless power at distances of 5-100m for indoor/outdoor charging of low-power devices; *2) Efficiency:* Boost the end-to-end power transfer efficiency (up to a fraction of percent/a few percent); *3) Non-line of sight (NLoS):* Support LoS and NLoS to widen the practical applications of this network; *4) Mobility support:* Support mobile receivers, at least for those at pedestrian speed; *5) Ubiquitous accessibility:* Support ubiquitous power accessibility within the network coverage area; *6) Seamless integration of wireless communication and wireless power:* Interoperate wireless communication and wireless power via a unified wireless information and power transfer (WIPT); *7) Safety and health:* Resolve the safety and health issues of RF systems and comply with the regulations; *8) Energy consumption:* Limit the energy consumption of the energy-constrained RF powered devices.

**Power Requirements and Consumption of Sensors and Devices**
The Integrated Circuit industry is moving from the traditional computing power paradigm towards a power efficiency (lowest joule per operation) paradigm. This ultra-low power (ULP) electronics has opened the door to numerous applications in sensor networks and IoT that do not need nm technology with billions of gates. Sensor nodes commonly require power for the sensor itself, the data processing circuitry and the wireless data link (e.g. a few bits/s for temperature sensors to a few kbits/s for ECG or blood pressure monitoring). The first two functions commonly require less power. This can be attributed to the fact that while CMOS technology scaling has conventionally provided exponential benefits for the size and power consumption of digital logic systems, analog RF components, necessary for the data link, have not seen a similar power scaling. In [Ay:2011], a CMOS image sensor consumes only 14.25µW. In [ADMP801], low power microphones consume 17µW and an ADC digitizing the microphone output consumes 33µW. Popular protocols for sensor networks include Zigbee and low power Bluetooth whose commercial-of-the-shelf transmitters consume 35mW [CC2541]. WiFi is more power-hungry. Despite the progresses in the WiFi industry to design chipsets for IoT applications by e.g. reducing power consumption in the standby mode to 20µW, an active WiFi transmission consumes around 600mW [Gainspan, CC3100MOD]. Nevertheless, in recent years, there has been significant enhancement with integrated ULP System on Chip (SoC) and duty-cycled radio whose power consumption is nowadays in the order of 10-100µW using custom protocols supporting 10-200kbps [Zhang:2013, Kim:2011, Verma:2010, Pandey:2011]. The use of passive WiFi is also an alternative to generate 802.11b transmission over distances of 10-30m (in line-of-sight and through walls) while only consuming 10 and 60µW for 1 and 11 Mbps transmissions, respectively (3 to 4 orders of magnitude lower than existing WiFi chipsets) [Kellogg:2016].

***Observation:*** 10-100µW is enough to power modern wireless sensors and low-power devices.

## WPT RF Design

Since Tesla's attempt in 1899, all WPT experiments in 1960-2000 were targeting long-distance and high power transmissions with applications such as Solar Powered Satellite and wireless-powered aircraft [Brown:1984]. More recently, there has been a significant interest in WPT and WEH for relatively low-power (e.g., from µW to a few W) delivery over moderate distances (e.g., a few m to hundreds of m) [Falkenstein:2012, Popovic:2013], owing to the fast-growing need to build reliable and convenient wireless power systems for remotely charging various low- to medium-power devices, such as RFID tags, wireless sensors, consumer electronics [Visser:2013, Popovic:2013a]. The interest in far-field wireless power has spurred the creation of initiatives like COST IC1301 [Carvalho:2014] and a small number of start-ups in recent years, namely Drayson Technologies, Powercast, Energous, Ossia.

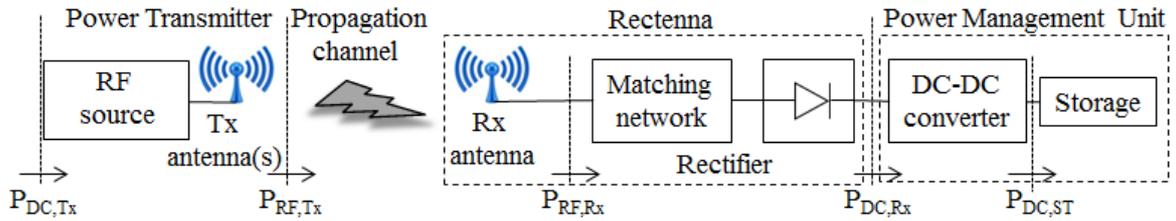

**Fig. 1**: Block diagram of a conventional far-field WPT architecture.

Fig. 1 shows a generic wireless power delivery system, which consists of an RF transmitter and an energy harvester made of a rectenna (antenna and rectifier) and a power management unit (PMU). Since the quasi-totality of electronics requires a DC power source, a rectifier is required to convert RF to DC. The recovered DC power then either supplies a low power device directly, or is stored in a battery or a super capacitor for high power low duty-cycle operations. It can also be managed by a DC-to-DC converter before being stored. Referring to Fig 1, the end-to-end power transfer efficiency $e$ can be expressed as

$$e = \frac{P_{DC,ST}}{P_{DC,Tx}} = \underbrace{\frac{P_{RF,Tx}}{P_{DC,Tx}}}_{e_1} \underbrace{\frac{P_{RF,Rx}}{P_{RF,Tx}}}_{e_2} \underbrace{\frac{P_{DC,Rx}}{P_{RF,Rx}}}_{e_3} \underbrace{\frac{P_{DC,ST}}{P_{DC,Rx}}}_{e_4}. \tag{1}$$

In WEH, the transmitter in Fig 1 is an RF communication transmitter, not controllable and not optimized for power delivery purposes. Given the typical power density between $10^{-3}$ and $10^{-1}$ µW/cm$^2$ observed indoor and outdoor at distances from 25 to 100m from a GSM900 base station, WEH is thought unlikely sufficient for powering devices with a few cm$^2$ in size requiring 10-100µW [Visser:2013]. In WPT, the power transmitter of Fig 1 can be fully optimized. Therefore, WPT offers more control of the design and room for enhancement of $e$. We briefly review the techniques used to enhance $e_1$, $e_2$, $e_3$ and $e_4$.

The <u>DC-to-RF conversion efficiency</u> $e_1$ maximization can leverage a rich literature on power amplifier (PA) design and rely on transmit signals with constrained Peak-to-Average Power Ratio (PAPR).

The RF-to-RF conversion efficiency $e_2$ is a bottleneck and requires highly directional transmission. Common approaches in the RF literature rely on real-time reconfiguration of time-modulated arrays based on localization of the power receivers [Masotti:2016], phased-arrays [Takahashi:2011] or retrodirective arrays [Miyamoto:2002].

The RF-to-DC conversion efficiency $e_3$ maximization relies on the design of efficient rectennas. A rectenna harvests electromagnetic energy, then rectifies and filters it using a low pass filter. Its analysis is challenging due to its nonlinearity, which in turn renders its implementation hard and subject to several losses due to threshold and reverse-breakdown voltage, devices parasitics, impedance matching, harmonic generation [Yoo:1992,Strassner:2013,Valenta:2014]. In WPT, the rectenna can be optimized for the specific operating frequencies and input power level. It is more challenging in WEH since the rectenna is designed for a broad range of input power densities (from a few nW/cm$^2$ to a few µW/cm$^2$) and spectrum (TV, WiFi, 2/3/4G) [Costanzo:2016]. In order to address the large aggregate frequency spectrum of ambient RF signals, multiband [Masotti:2013,Pinuela:2013, Niotaki:2014,Belo:2016] and broadband [Kimionis:2017,Song:2015,Sakaki:2014] rectifier designs have been proposed. In the case of multiband designs one may maximize $e_3$ over a number of narrowband frequency regions, whereas in the case of broadband (ultra-wideband) designs one may cover a much larger frequency band however sacrificing the obtained maximum efficiency. Various rectifier technologies exist, including the popular Schottky diodes [Falkenstein:2012,Hagerty:2004], CMOS [Le:2008], active rectification [Roberg:2012], spindiode [Hemour:2014], backward tunnel diodes [Lorenz:2015]. Assuming $P_{RF,Tx}$ =1W, 5-dBi Tx/Rx antenna gain, a continuous wave (CW) at 915MHz, $e_3$ of state-of-the-art rectifiers is 50% at 1m, 25% at 10m and about 5% at 30m [Hemour:2014]. This severely limits the range of WPT. Moreover, with the current rectifier technologies, $e_3$ drops from 80% at 10mW to 40% at 100µW, 20% at 10µW and 2% at 1µW [Valenta:2014, Hemour:2014]. This is due to the diode not being easily turned on at low input power. Enhancements for the very low power regime (below 1 µW) rely on spindiodes [Hemour:2014] and backward tunnel diodes [Lorenz:2015]. For typical input power between 1 µW and 1mW, low barrier Schottky diodes remain the most competitive and popular technology [Hemour:2014,Costanzo:2016, Valenta:2014]. $e_3$ also decreases as the frequency increases due to parasitic losses [Valenta:2014]. The rectifier topology also impacts $e_3$. A single diode is preferred at low power (1-500µW) and multiple diodes (voltage doubler/diode bridge/charge pump) favoured above 500µW [Costanzo:2016, Boaventura:2013]. The efficiency is also dependent on the input power level and the output load variations. One possibility to minimize sensitivity to output load variation is to use a resistance compression network [Niotaki:2014], while topologies using multiple rectifying devices each one optimized for a different range of input power levels can enlarge the operating range versus input power variations and avoid, within the power range of interest, the saturation effect (that creates a sharp decrease in $e_3$) induced by the diode breakdown [Sun:2013]. This can be achieved using e.g. a single-diode rectifier at low input power and multiple diodes rectifier at higher power.

Interestingly the rectenna design is not the only factor influencing $e_3$. Due to the rectifier nonlinearity, the input waveform (power and shape) also influences $e_3$ in the low input power regime (1µW-1mW) [Trotter:2009, Boaventura:2011, Valenta:2013,Valenta:2015, Collado:2014]. A 20dB gain (in terms of $P_{DC,Rx}$) of a multisine over a CW excitation at an average input power of -15dBm was shown in [Valenta:2013]. It is to be noted though that the output filter is also important in relation to the tone separation in order to boost the performance of multisine waveform [Boaventura:2014,Pan:2015]. High PAPR signals were also shown beneficial in [Collado:2014]. It was nevertheless argued in [Blanco:2016] that the instantaneous power variance is more accurate than PAPR to characterise the effect of modulation on the rectifier efficiency. Suitable signals and waveforms therefore exploit the nonlinearity to boost $e_3$ at low input powers and extend WPT range [Boaventura:2013]. Modulation also has an impact on $e_3$. In [Vera:2010], QPSK modulation was shown to be beneficial to $e_3$ compared to a CW in the low power regime -20dBm to 0dBm. [Fukuda:2014,Sakaki:2014] reported somewhat contradicting behaviors in the higher input power regime of 0-20dBm. In [Fukuda:2014], PSK and QAM modulations were shown beneficial to $e_3$ compared to a CW, while they were shown detrimental in [Sakaki:2014]. Finally, [Bolos:2016] argues that one may or may not get an advantage from using multisines or other modulated signals depending on the load and input power.

The DC-to-DC conversion efficiency $e_4$ is enhanced by dynamically tracking the rectifier optimum load, e.g. dc-to-dc switching converters dynamically track the maximum power point (MPP) condition [Dolgov:2010,Costanzo:2012]. Due to the variable load on the rectenna, the changes in diode impedance with power level and the rectifier nonlinearity, the input impedance of the rectifier becomes highly variable, which renders the matching hard, not to mention a joint optimization of the matching and load for multisine signals [Bolos:2016].

Interestingly, the maximization of e is not achieved by maximizing $e_1$, $e_2$, $e_3$, $e_4$ independently from each other, and therefore simply concatenating the above techniques. This is because $e_1$, $e_2$, $e_3$, $e_4$ are coupled with each other due to the rectifier nonlinearity, especially at input power range 1 µW -1 mW. Indeed, $e_3$ is a function of the input signal shape and power to the rectifier and therefore a function of 1) the transmit signal (beamformer, waveform, power allocation) and 2) the wireless channel state. Similarly, $e_2$ depends on the transmit signal and the channel state and so is $e_1$, since it is a function of the transmit signal PAPR.

Some recent approaches optimize the system using numerical software tools based on a combination of full-wave analysis and nonlinear harmonic balance techniques in order to account for nonlinearities and electromagnetic couplings [Masotti:2016,Costanzo:2014]. This approach would provide very high accuracy but has the drawback to hold only for an offline optimization of the system, not for an adaptive WPT whose transmit signal is adapted every few ms as a function of the channel state, not to mention for an entire WPT network with multiple transmitters and receivers.

***Observations:***
**First**, the majority of the technical efforts in the wireless power literature has been devoted to the design of the energy harvester but much less emphasis has been put on signals design for WPT.
**Second**, the emphasis has remained on point-to-point (single user) transmission.
**Third**, research has recognized the importance of non-linearity of the rectenna in WPT system design but has focused to a great extent on decoupling the WPT design by optimizing the transmitter and the energy harvester independently from each other.
**Fourth**, multipath and fast fading, critical in NLoS, have been ignored despite playing a key role in wireless transmissions and having a huge impact on the signal shape and power at the rectenna input. Recall indeed that multipath has for consequences that transmit and received (at the input of the rectenna) waveforms are completely different.
**Fifth**, WPT design has remained very much centered around an open-loop approach with waveform being static and beamforming relying on tags localization, not on the channel state.
**Sixth**, the design of the transmit signals is heuristic (with conclusions exclusively based on observations from measurements using various predefined and standard waveforms) and there exist no systematic approach and performance bounds to design and evaluate them. Waveform and beamformer have been studied independently, despite being part of the same transmit signal.

**To tackle the listed challenges, we need:**
**First**, a closed-loop and adaptive WPT architecture with a reverse communication link from the receiver to the transmitter that is used to support various functions such as channel feedback/training, energy feedback, charging control, etc. The transmitter should be able to flexibly adjust the transmission strategy jointly optimized across space and frequency (through beamforming and waveform) in accordance with the channel status (commonly called Channel State Information - CSI), and thus, renders state-of-the-art Multi-Input Multi-Output (MIMO) processing an indispensable part of WPT. An example of a closed-loop and adaptive WPT architecture is displayed in Fig. 2.

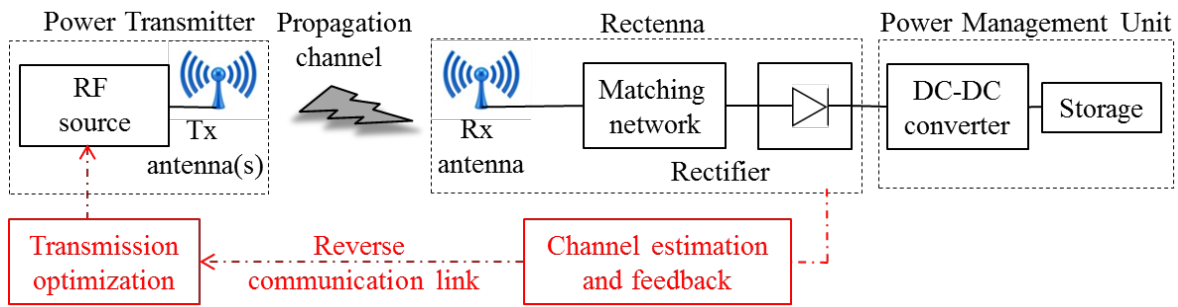

**Fig. 2**: Block diagram of a closed-loop and adaptive WPT architecture.

**Second**, a systematic approach to design and optimize, as a function of the channel, the signal at the transmitter (encompassing beamforming and waveform) so as to maximize $e_2 \times e_3$ subject to transmit power and PAPR constraints. This requires to capture the rectifier nonlinearity as part of the signal design and optimization. Such a systematic design methodology will lead to the implementation of efficient strategies as part of the "transmission optimization" module of Fig. 2.

**Third**, a link and system design approach that takes wireless power from a rectenna paradigm to a network paradigm with multiple transmitters and/or receivers. Instances of such network architectures may be a deployment of co-located transmit antennas delivering power to multiple receivers or the dense and distributed deployment of well-coordinated antennas/transmitters, as illustrated in Fig. 3.

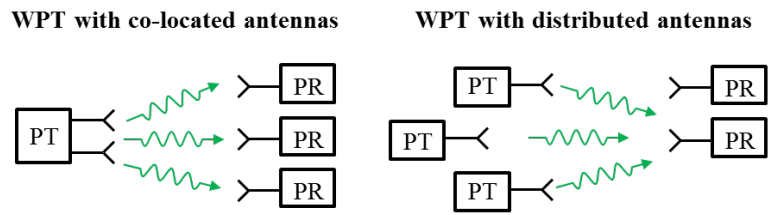

**Fig. 3**: Illustration of a WPT network with co-located/distributed transmit antennas and multiple receivers (P T/R: Power Transmitter/Receiver).

## Leveraging Ideas from Wireless Communications

The fundamental limits of a communication network design lie in information and communication theories that derive the capacity of wireless channels (point-to-point, broadcast, multiple access, interference channel with single and multiple antennas) and identify transmission and reception strategies to achieve it, most commonly under the assumption of a linear communication channel with additive white Gaussian noise (AWGN) [68-70]. In the 70s till early 2000s, the research emphasis was on a link optimization, i.e. maximizing the point-to-point spectral efficiency (bits/s/Hz) with advances in modulation and waveforms, coding, MIMO, Channel State Information (CSI) feedback and link adaptation, communication over (multipath) fading channel. CSI feedback enables to dynamically adapt the transmission strategies as a function of the channel state. It leads to a drastic increase in rate and complexity reduction in receiver design. The emphasis in 4G design shifted towards a system optimization, with a more interference-centric system design. MIMO evolved into a multi-link/user/cell MIMO. Multiple users are scheduled in the same time-frequency resource onto (ideally) non-interfering spatial beams. This led to significant features such as multi-user MIMO, multi-user fairness and scheduling and multi-point cooperation. The availability of accurate CSI at the Transmitter (CSIT) is also crucial for multi-user multi-antenna wireless communication networks, for beamforming and interference management purposes. Some promising technologies consist in densifying the network by adding more antennas either in a distributed or in a co-localized manner. The distributed deployment leads to dense network (with a high capacity backbone) requiring interference mitigation techniques, commonly denoted as Coordinated Multi-Point transmission and reception (CoMP) in 3GPP, and classified into joint processing (or Network MIMO) and coordinated scheduling, beamforming and power control. Co-localized deployment leads to Massive/Large-Scale MIMO where a base station designs pencil beams (with large beamforming gain) serving its own users, using per-cell design rules, while simultaneously avoiding inter-cell interference. The reader is invited to consult [Clerckx:2013] for fundamentals and designs of state-of-the-art MIMO wireless communication networks.

*Observations:*
**First**, wireless power and communication systems share the same medium and techniques inspired by communications, such as MIMO, closed-loop operation, CSI acquisition and transmitter coordination are expected to be useful to WPT.
**Second**, existing techniques developed for wireless communications cannot be directly applied to wireless power, due to their distinct design objectives (rate vs energy), practical limitations (hardware and power constraints), receiver sensitivities (e.g. -30dBm for rectenna vs -60dBm for information receivers), interference (beneficial in terms of energy harvesting vs detrimental in communications) and models (linear wireless communication channel vs nonlinear wireless power channel due to the rectifier).

**WPT Signal and System Design**

Aside the traditional WPT RF design, a new and complementary line of research on communications and signal design for WPT has emerged recently in the communication literature [Zeng:2017] and is briefly reviewed in the sequel. This includes among others the design of efficient transmit signals (including waveform, beamforming and power allocation), CSI acquisition strategies, multiuser transmission strategies, integration with communications and system prototyping. Importantly, the nonlinearity of the rectifier has to be captured as part of the signal and system design and optimization as it induces coupling among the various efficiencies.

Let us first consider a point-to-point scenario with a single transmitter and receiver. The first systematic approach towards signal design in adaptive closed-loop WPT was proposed in [Clerckx:2015,Clerckx:2016], where the transmit signal, accounting jointly for multisine waveform, beamforming and power allocation, is optimized as a function of the CSI to maximize $e_2 \times e_3$ subject to optional transmit PAPR constraints. Uniquely, such a signal design resolves some limitations of the WPT literature by optimally exploiting a beamforming gain, a frequency diversity gain and the rectifier nonlinearity. The rectifier nonlinearity was modelled using a Taylor expansion of the diode characteristic, which is a popular model in the RF literature [Boaventura:2011,Boaventura:2013]. The phases of the optimized waveform can be computed in closed-form while the magnitudes result from a non-convex optimization problem that can be solved using convex optimization techniques, so-called Reversed Geometric Program (GP). Multiple observations were made in [Clerckx:2015,Clerckx:2016]. *First*, it was observed that the derived adaptive and optimized signals designed accounting for the nonlinearity are more efficient than non-adaptive and non-optimized multisine signals (as used in [Trotter:2009, Boaventura:2011, Valenta:2013, Valenta:2015]). *Second*, the rectifier nonlinearity was shown essential to design efficient wireless power signals and ignoring it leads to inefficient signal design in the low power regime. *Third*, the optimized waveform design favours a power allocation over multiple frequencies and those with stronger frequency-domain channel gains are allocated more power. *Fourth*, multipath and frequency-selective channels were shown to have significant impact of DC output power and waveform design. Though multipath is detrimental to performance with non-adaptive waveforms, it is beneficial with channel-adaptive waveform and leads to a frequency diversity gain.

As an illustration, Fig. 4 (top) displays the magnitude of the frequency response of a given realization of the wireless channel over a 10MHz-bandwidth. We consider a multisine waveform with 16 sinewaves uniformly spread within the 10MHz. Assuming this channel has been acquired to the transmitter, the magnitudes of the optimized waveform on the 16 frequencies can be computed and are displayed on Fig. 4 (bottom). Interestingly, in contrast with the waveforms commonly used in the RF literature [Trotter:2009, Boaventura:2011, Valenta:2013, Valenta:2015, Collado:2014] that are non-adaptive to the channel state, the optimized adaptive waveform has a tendency to allocate more power to frequencies exhibiting larger channel gains.

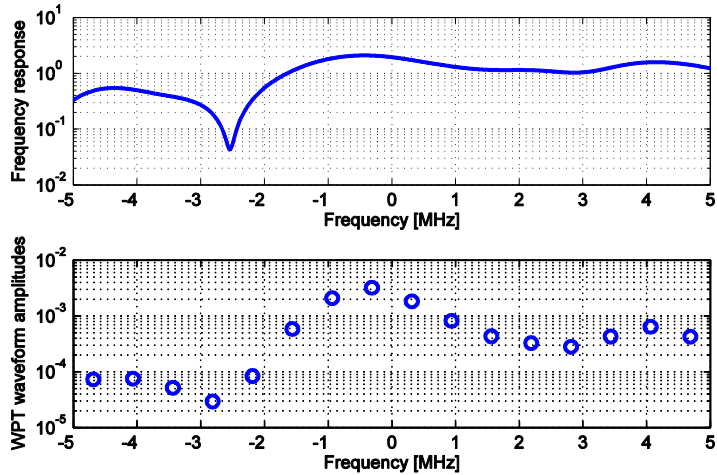

**Fig. 4**: Frequency response of the wireless channel and WPT waveform magnitudes (N = 16) for 10 MHz bandwidth [Clerckx:2016].

The performance benefits of those optimized channel-adaptive multisine waveforms over the non-adaptive design (in-phase multisine with uniform power allocation) of [Trotter:2009, Boaventura:2011, Valenta:2013, Valenta:2015] has been validated using ADS and PSpice simulations with a single series rectifier in a WiFi-like environment at 5.18GHz for an average input power of about -12dBm in [Clerckx:2015] and -20dBm in [Clerckx:2016, Clerckx:2017]. As illustrated in Fig. 5, for a single transmit antenna and a single series rectifier subject to an average input power of -12dBm and multipath fading, the gains in terms of harvested DC power are very significant with over 100% gains for 4 sinewaves and about 200% gain for 8 sinewaves over the non-adaptive design. Significant performance gains have also been validated in [Clerckx:2016] at -20dBm average input power for various bandwidths and in the presence of multiple transmit antennas where waveform and beamforming are jointly designed. Moreover, it was interestingly shown in [Clerckx:2017] that the systematic signal design approach of [Clerckx:2016] actually is applicable to and provides gains (100%-200%) in a wide range of rectifier topologies, e.g. single series, voltage doubler, diode bridge. Details on circuit design and simulation assumptions can be found in [Clerckx:2016, Clerckx:2017].

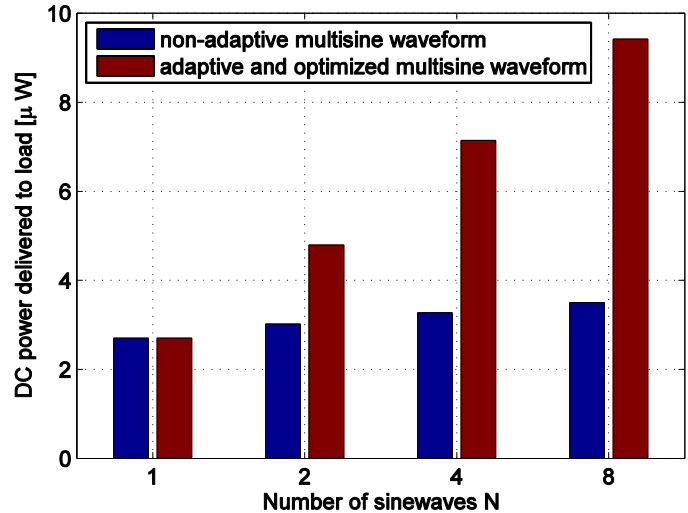

**Fig. 5**: DC power vs. number of sinewaves N for adaptive and non-adaptive waveforms.

Importantly, systematic and optimized signal designs of [Clerckx:2016] also show that contrary to what is claimed in [Valenta:2015,Collado:2014], maximizing PAPR is not always the right approach to design efficient wireless power signal. High PAPR is a valid metric in WPT with multisine waveforms if the channel is frequency flat, not in the presence of multipath and frequency selectivity. This can be inferred from Fig 5 where the non-adaptive multisine waveform leads to a much lower DC power despite exhibiting a significantly higher transmit PAPR compared to the adaptive waveform. Recall indeed that the adaptive waveform will unlikely allocate power uniformly across all sinewaves, as it emphasizes the ones corresponding to strong frequency domain channel. This leads to waveforms whose PAPR is lower than the non-adaptive in-phase multisine waveform with uniform power allocation.

Results in [Clerckx:2016] also highlighted the potential of a <u>large scale multi-sine multi-antenna closed-loop WPT architecture</u>. In [Huang:2016,Huang:2017], such a promising architecture was designed and shown to be an essential technique in enhancing *e* and increasing the range of WPT for low-power devices. It enables highly efficient very far-field wireless charging by jointly optimizing transmit signals over a large number of frequency components and transmit antennas, therefore combining the benefits of pencil beams and waveform design to exploit the large beamforming gain of the transmit antenna array and the non-linearity of the rectifier at long distances. The challenge is on the large dimensional problem, which calls for a reformulation of the optimization problem. The new design enables orders of magnitude complexity reduction in signal design compared to the Reverse GP approach. Another low-complexity adaptive waveform design approach expressed in closed-form (hence, suitable for practical implementation) has been proposed in [Clerckx:2017] and shown to perform close to the optimal design. Fig 6 illustrates how the rectifier output voltage decreases with the range for several values of the number of sinewaves N and transmit antennas M in the multisine transmit waveform. By increasing both N and M, the range is expanded thanks to the optimized channel-adaptive multisine waveforms that jointly exploit a beamforming gain, a frequency diversity gain and the rectifier nonlinearity.

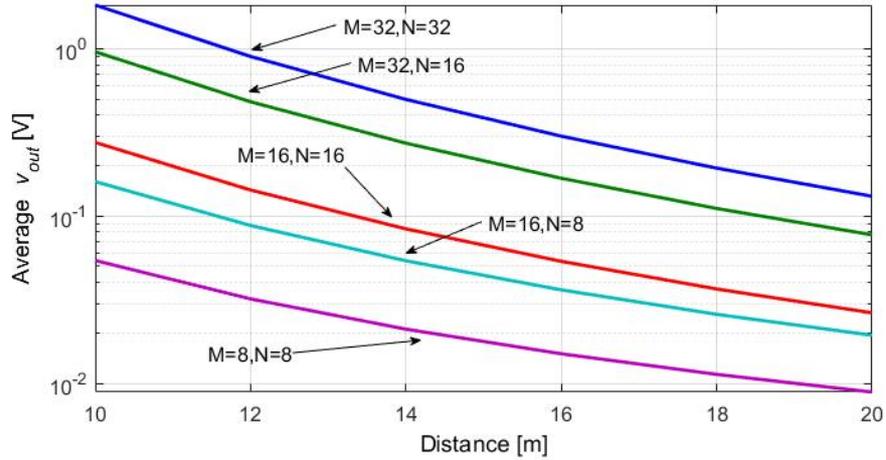

**Fig. 6**: Rectifier average output voltage as a function of the Tx-Rx distance [Huang:2017].

Note that in those recent progress on signal design, despite the presence of many transmit antennas and sinewaves, a single receive antenna and rectifier per terminal has been assumed. It would be interesting to understand how to extend the signal design to multiple receive antennas. This brings the problem of RF or dc combining or mixed RF-dc combining [Popovic:2014, Gutmann:1979, Shinohara:1998].

Discussions so far assumed deterministic multisine waveforms. It is of significant interest to understand how modulated waveforms perform in comparison to deterministic waveforms and how modulation could be tailored specifically for WPT to boost the end-to-end power transfer efficiency. This would also open the way to understanding how to design unified and efficient signals for the simultaneous transmission of information and power. A modulated waveform exhibits randomness and this randomness has an impact on the amount of harvested DC power. Interestingly, it was shown in [Clerckx:2017b] that for single-carrier transmission, modulation using circularly symmetric complex Gaussian (CSCG) inputs is beneficial to the performance compared to an unmodulated continuous wave. This gain comes from the large fourth order moment offered by CSCG inputs which is exploited by the rectifier nonlinearity. Even further gain can be obtained using asymmetric Gaussian inputs [Varasteh:2017]. On the other hand, for multi-carrier transmission, modulation using CSCG inputs was shown in [Clerckx:2017b] to be less efficient than a multisine because of the independent randomness across carriers, which leads to random fluctuations. This contrasts with the periodic behavior of deterministic multisine waveforms that are more suitable to turn on and off the rectifier periodically. Interestingly, it was noted in [Clerckx:2017b] that PAPR is not an appropriate metric to assess the suitability of a general modulated waveform for WPT. Nevertheless, despite all those recent progress on signal design for WPT, the optimum input distribution remains unknown and so is the optimum waveform.

We now understand that a systematic design of waveform design (including modulation, beamforming, power allocation) is a key technique to jointly exploit a beamforming gain, the channel frequency-selectivity and the rectifier nonlinearity, so as to enhance the end-to-end power transfer efficiency and the range of WPT. One challenge is that those waveforms have been designed assuming perfect CSI at the transmitter. In practice, this is not the case and the transmitter should find ways to acquire the CSI. Various strategies exist, including forward-link training with CSI feedback, reverse-link training via channel reciprocity, power probing with limited feedback [Zeng:2017]. The first two are reminiscent of strategies used in modern communication systems [Clerckx:2013]. The last one is more promising and tailored to WPT as it is implementable with very low communication and signal processing requirements at the terminal. It relies on harvested DC power measurement and on a limited number of feedback bits for waveform selection and refinement [Huang:2017b]. In the waveform selection strategy, the transmitter transmits over multiple timeslots with every time a different waveform precoder within a codebook, and the receiver reports the index of the precoder in the codebook that leads to the largest harvested energy. In the waveform refinement strategy, the transmitter sequentially transmits two waveforms in each stage, and the receiver reports one feedback bit indicating an increase/decrease in the harvested energy during this stage. Based on multiple one-bit feedback, the transmitter successively refines waveform precoders in a tree-structured codebook over multiple stages.

Wireless power networks are however not limited to a single transmitter and receiver. Let us now consider the presence of a single transmitter and multiple users/receivers, with each receiver equipped with one rectenna. In this multi-user deployment, the energy harvested by a given rectenna in general depends on the energy harvested by the other rectennas. Indeed, a given waveform may be suitable for a given rectenna but found inefficient for another rectenna. Hence, there exists a trade-off between the energy harvested by the different rectennas. The energy region formulates this trade-off by expressing the set of all rectenna harvested energy that are simultaneously achievable. It is mathematically written as a weighted sum of harvested energy where by changing the weights we can operate on a different point of the energy region boundary. Strategies to design WPT waveforms in this multi-user/rectenna deployment were discussed in [Clerckx:2016, Huang:2017]. Fig 7 illustrates such an energy region for a two-user scenarios with a multisine waveform spanning 20 transmit antennas and 10 frequencies. The key message here is that by optimizing the waveform to jointly deliver power to the two users simultaneously, we get an energy region ('weighted sum') that is larger than the one achieved by doing a timesharing approach, i.e. TDMA, where the transmit waveform is optimized for a single user at a time and each user is scheduled to receive energy during a fraction of the time.

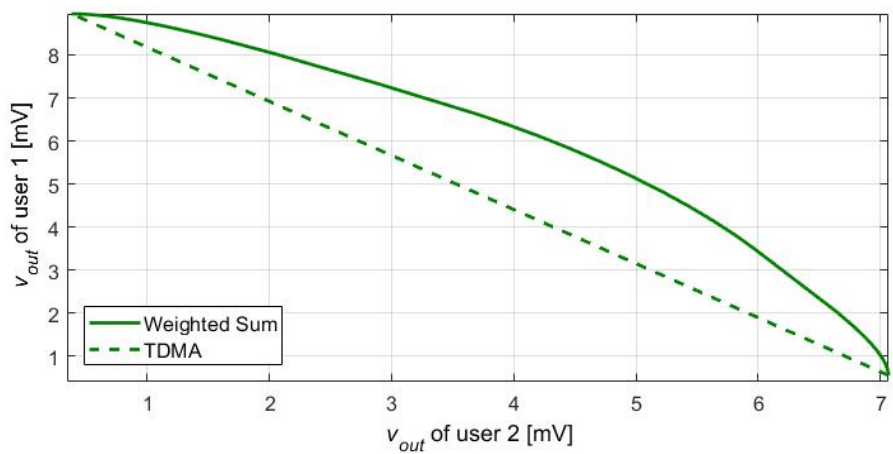

**Fig. 7**: 2-user energy region with M = 20 and N = 10 [Huang:2017].

Moving towards an entire network made of many transmitters and receivers, a network architecture needs to be defined [Zeng:2017]. This may consist in having all transmitters cooperating to jointly design the transmit signals to multiple receivers or having a local coordination among the transmitters such that a given receiver is served by a subset of the transmitters or the simplest scenario where each receiver is served by a single transmitter. This leads to different resource allocation and charging control strategies (centralized vs distributed) and requirements in terms of CSI sharing and acquisition at the different transmitters. Results in [Zeng:2017] show that distributing antennas in a coverage area (as in Fig 3) and enabling cooperation among them distributes energy more evenly in space and therefore potentially enhances the ubiquitous accessibility of wireless power, compared to a co-located deployment. It also avoids creating strong energy beams in the direction of the users, which is desirable from a health and safety perspective.

Demonstrating the feasibility of the aforementioned signal and system designs through prototyping and experimentation remains a largely open challenge. It requires the implementation of a closed-loop WPT architecture with a real-time over-the-air transmission based on a frame structure switching between a channel acquisition phase and wireless power transfer phase. The channel acquisition needs to be done at the millisecond level (similarly to CSI acquisition in communication). Different blocks need to be built, namely channel estimation, channel-adaptive waveform design and rectenna. The first prototype of a closed-loop WPT architecture based on channel-adaptive waveform optimization and dynamic channel acquisition, as illustrated in Fig 8, was recently reported in [Kim:2017] with further enhancements in [Kim:2017b]. Importantly, all experimental results validate the theory developed in [Clerckx:2016,Clerckx:2017] and fully confirm the following observations: 1) diode nonlinearity is beneficial to WPT performance and is to be exploited in systematic waveform design, 2) the wireless propagation channel has a significant impact on signal design and system performance, 3) CSI acquisition and channel-adaptive waveforms are essential to boost the performance in frequency-selective channels (as in NLoS scenarios), 4) larger bandwidths benefit from a channel frequency diversity gain, 5) PAPR is not an accurate metric to assess and design waveforms for WPT in general frequency-selective channels. The performance gain of channel-adaptive multisine waveforms versus non-adaptive multisine waveform in a NLoS deployment with a single antenna at the transmitter and receiver is illustrated in Fig 9. We note the significant boost of the average harvested DC power at the rectenna output by 105% over an open-loop WPT architecture with non-adaptive multisine waveform (with the same number of sinewaves) and by 170% over a continuous wave.

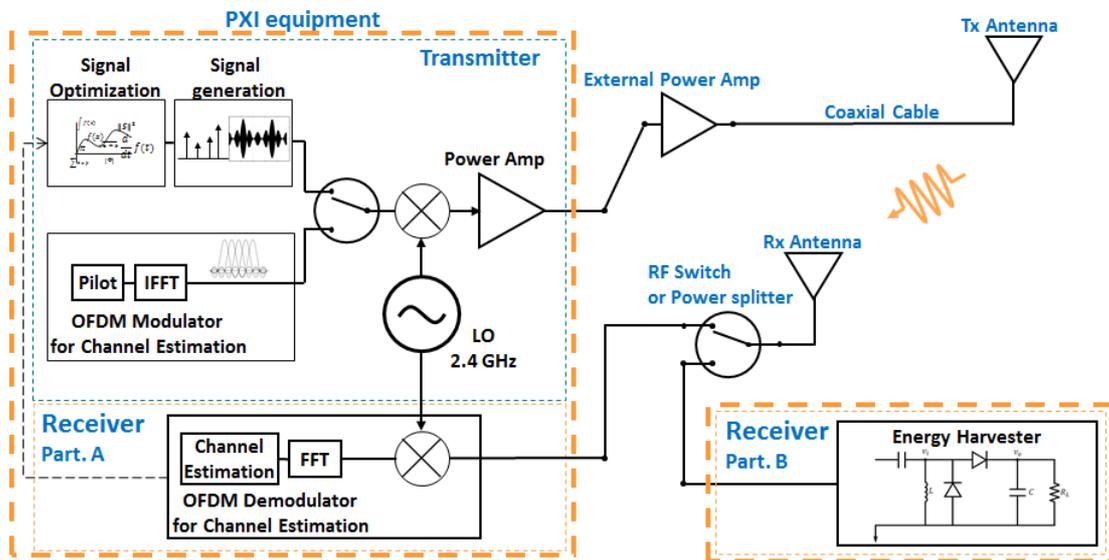

**Fig. 8**: Prototype architecture with 3 key modules: signal optimization, channel acquisition and energy harvester [Kim:2017,Kim:2017b].

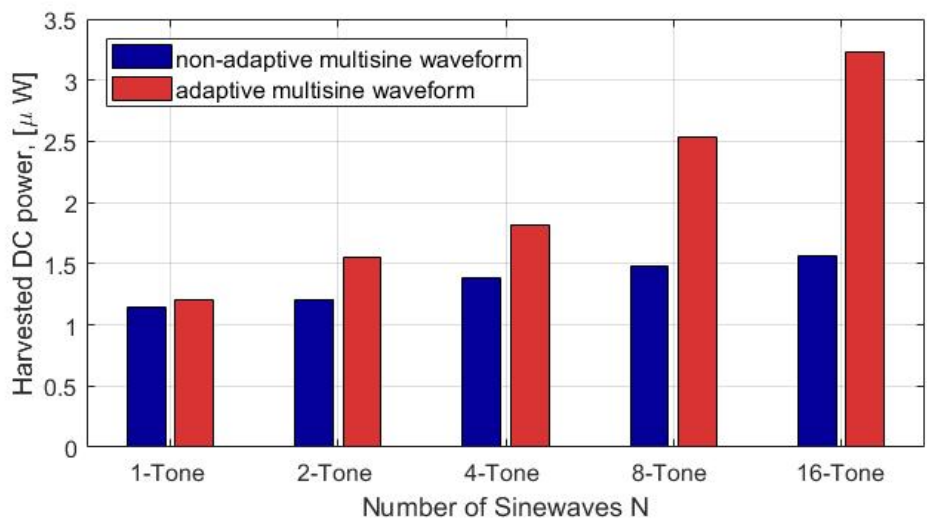

**Fig. 9**: Harvested DC power with the architecture of Fig 8 in an indoor NLoS deployment as a function of N uniformly spread within a 10MHz bandwidth [Kim:2017b].

Ultimately wireless power and wireless communications will have to be integrated. This calls for a unified Wireless Information and Power Transfer (WIPT) paradigm. A major challenge is to characterize the fundamental tradeoff between conveying information and energy wirelessly [Varshney:2008, Grover:2010, Zhang:2013] and to identify corresponding transmission strategies. Leveraging the aforementioned wireless power signal designs, it has been shown that the rectifier nonlinearity has profound impact on the design of WIPT [Clerckx:2016b,Clerckx:2017b, Varasteh:2017]. In contrast with the classical capacity achieving CSCG input distribution, rectifier nonlinearity leads to input distribution that are asymmetric Gaussian in single-carrier transmission over frequency flat channels [Varasteh:2017] and to non-zero mean Gaussian in multi-carrier transmissions [Clerckx:2016b,Clerckx:2017b]. Nevertheless, the optimal input distribution and transmit signal strategy for WIPT remain unknown. Those refreshing results are in sharp contrast with earlier results of [Varshney:2008, Grover:2010, Zhang:2013] that ignore the rectifier nonlinearity and therefore rely on the conventional capacity-achieving CSCG input distribution.

*Observations:*
**First**, the above results show the huge potential in a systematic signal and system design and optimization approach towards efficient WPT and WIPT that accounts for the unique characteristic of wireless power, namely the non-linearity.
**Second**, the nonlinearity radically changes the design of WPT and WIPT: it 1) leads to a WIPT design different from that of conventional wireless communication (whose channel is assumed linear), 2) favours a different input distribution, signal design, transceiver architecture and use of the RF spectrum, 3) is beneficial to increase the rectifier output DC power and enlarge the rate-energy region.
**Third**, an adaptive signal design approach provides a different paradigm compared to the traditional WPT design. It leads to an architecture where the rectenna is fixed as much as possible (e.g. with a fixed load) but the transmit signal is adaptive in contrast with the approach in the RF literature where the waveform is fixed and the rectenna/PMU is adaptive (e.g. dynamic load control). Since the wireless channel changes quickly (10ms order), it can be impractical for energy-constrained devices to dynamically compute and adjust the matching and the load as a function of the channel. Even though both approaches are complementary, the adaptive signal approach makes the transmitter smarter and decreases the need for power-hungry optimization at the devices. Nevertheless, adaptation implies acquiring CSIT, which is an important challenge to be addressed. Ultimately, it is envisioned that an entire end-to-end optimization of the system should be conducted, likely resulting in an architecture where the transmit signals and the rectennas adapt themselves dynamically as a function of the channel state.

*Conclusions:*

An integrated signal and system optimization has been introduced as the strategic approach to realize the first generation of a mobile power network and to enable energy autonomy of pervasive devices, such as smart objects, sensors and embedded systems in a wide range of operating conditions.
It has been shown that the nonlinear nature of this design problem, both for the transmitter and for the receiver sides, must be accounted for the signal and the circuit- level design. In this way, a new architecture of the system is foreseen enabling simultaneous WPT and WIPT while enhancing the power transfer efficiency at ultra-low power levels. Techniques for dynamic tracking of the channel changes need to be exploited to adaptively modify the transmitted energy, both in terms of its waveform shape and of its intensity, with the twofold advantage of reducing the complexity of the rectenna and of the PMU design while keeping the rectenna itself in its own optimum operating conditions.